\documentclass[preprint]{elsarticle}

\usepackage{lineno,hyperref}
\modulolinenumbers[5]

\usepackage{amsfonts,bbm,amssymb,amsmath}
\usepackage{dsfont}
\usepackage{txfonts}
\usepackage{hyperref}
\usepackage{xspace}
\usepackage{mathrsfs}
\usepackage{graphicx}
\usepackage[toc,page]{appendix}

\def\bra#1{\mathinner{\langle{#1}|}}
\def\ket#1{\mathinner{|{#1}\rangle}}
\def\expect#1{\langle#1\rangle}

\makeatletter
\def\maketag@@@#1{\hbox{\m@th\normalfont\normalsize#1}}
\makeatother

\makeatletter
  \def\my@tag@font{\normalsize}
  \def\maketag@@@#1{\hbox{\m@th\normalfont\my@tag@font#1}}
  \let\amsmath@eqref\eqref
  \renewcommand\eqref[1]{{\let\my@tag@font\relax\amsmath@eqref{#1}}}
\makeatother

\renewcommand{\d}[1]{\ensuremath{\operatorname{d}\!{#1}}}
\newcommand{\uq}{\ensuremath{U_q(\mathfrak{sl}_2)}\xspace}
\newcommand{\uLq}{\ensuremath{U_q(L\mathfrak{sl}_2)}\xspace}

\newcommand{\ii}{ {\rm i} }
\newcommand{\dd}{ \dd }

\newcommand{\ZZ}{\mathbb{Z}}

\newcommand{\CC}{\mathbb{C}}
\newcommand{\y}{{\rm y}}
\newcommand{\x}{{\rm x}}
\newcommand{\z}{{\rm z}}

\newcommand{\mm}[1]{{\mathbf{#1}}}
\newcommand{\Hp}{\mathcal{H}}
\newcommand{\Va}{\mathcal{V}}
\def\z{{\rm z}}
\def\x{{\rm x}}
\def\y{{\rm y}}
\def\tr{{{\rm tr}}}

\def\End{{\,{\rm End}\,}}
\def\one{\mathbbm{1}}
\def\Re{{\,{\rm Re}\,}}

\journal{Nuclear Physics B}

\begin{document}

\begin{frontmatter}
 
\title{Quasilocal conservation laws from semicyclic irreducible representations of \uq in $XXZ$ spin-$1/2$ chains}

\author{Lenart Zadnik, Marko Medenjak, and Toma\v{z} Prosen}

\address{Department of Physics, Faculty of Mathematics and Physics, University of Ljubljana, Jadranska 19, SI-1000 Ljubljana, Slovenia}

\begin{abstract}
We construct quasilocal conserved charges in the gapless ($|\Delta| \le 1$) regime of the Heisenberg $XXZ$ spin-$1/2$ chain, using semicyclic irreducible representations of \uq. 
These representations are characterized by a periodic action of ladder operators, which act as generators of the aforementioned algebra.
Unlike previously constructed conserved charges, the new ones do not preserve magnetization, i.e. they do not possess the $U(1)$ symmetry of the Hamiltonian. 
The possibility of application in relaxation dynamics resulting from $U(1)$-breaking quantum quenches is discussed.
\end{abstract}

\end{frontmatter}

\section{Introduction}
\label{intro}

In this paper we shall consider the anisotropic Heisenberg $XXZ$ spin-$1/2$ chain with periodic boundary conditions, from the point of view of theory of integrability. Heisenberg spin chains, providing successful theoretical description of magnetism-related phenomena in spin-chain materials \cite{spinchains}, can be studied analytically using the so called \emph{algebraic Bethe ansatz} method \cite{fad,kbi93}, resulting in an infinite family of conservation laws. These can then be applied in numerous ways to treat various aspects of equilibrium and nonequilibrium statistical physics of the model in consideration. An example of application of these results out-of-equilibrium is a rigorous derivation and evaluation \cite{pro11,enej,pro,proenej} of Mazur-Suzuki lower bound \cite{mazur,suzuki} for spin transport \cite{zotos}. The result of this evaluation in linear response theory implies strict ballistic property of high temperature spin transport in the thermodynamic limit in the $|\Delta| < 1$ regime of the anisotropic Heisenberg spin-$1/2$ Hamiltonian. This issue has been controversial in view of the fact that the {\em thermodynamic Bethe ansatz} approach to quantum spin transport allowed for different, mutually inconsistent results \cite{zotos1,klumper}. Another, closely related application of (quasi)local conserved operators is in relaxation dynamics that follows quantum quenches of integrable systems, where the precise formulation of the so-called {\em generalized Gibbs ensemble} is currently under intense investigation \cite{amsterdam,budapest,coop}.

The success of the algebraic Bethe ansatz method, in case of Heisenberg spin chains relies heavily on the existence of quantum group \uq and its universal $R$-matrix, satisfying the so called \emph{quantum Yang-Baxter equation} \cite{pasquier}. 
Using the fundamental, two-dimensional representations of these objects, Faddeev and his Leningrad school \cite{fad} developed a general technique which generates the Heisenberg Hamiltonian, together with the full family of local conserved charges in involution,
via the logarithmic derivatives of the quantum transfer matrix. However, in a recent progress \cite{pro,proenej,pereira}, other highest weight irreducible representations at the root-of-unity values of $q$, which densely populate the entire critical interval $-1 < \Delta < 1$, have been implemented to construct quasi-local conserved quantities relevant for quantum spin transport and quantum relaxation \cite{marcin}.
In the present paper, these constructions are generalized and extended, using \emph{semicyclic} irreducible representations \cite{kassel,arnaudon}. Here the highest and the lowest weight vectors are coupled by the periodic action of \uq-generators.

In the second section we briefly review the model and the structure of cyclic and semicyclic irreducible representations of \uq, as given (up to module isomorphism) in Refs. \cite{kassel} and \cite{arnaudon}. Also, the problem of periodicity of the generator action is reviewed: the fundamental commutation relations described in \cite{fad}, implying the conservation and involution of transfer-matrix related quantities, fail to hold in case of irreducible even-dimensional and certain odd-dimensional (semi)cyclic representations \cite{korff}. Since the anisotropy parameter is linked to the parameter $q$ -- the parameter of deformation of the quantum group and thus to the dimension of a representation, this imposes a restriction on the values of anisotropy parameter, for which this construction is valid.

In the third section we construct new quasilocal conserved quantities \eqref{monodromyderivative}, from valid odd-dimensional irreducible representations, using the formal procedure described in \cite{pro}. Quasilocality follows from the argument stated in the latter paper. As a direct consequence of periodicity of generator actions, the total magnetization in $z$ direction is not conserved by the newly constructed conserved operators. This is compatible with abundant degeneracies found in the spectrum of the $XXZ$ model at the root of unity anisotropies as a consequence of the loop algebra symmetry \cite{mccoy, korff2}. 

Finally, we propose in section 4 some potentially interesting applications of new quasilocal conserved operators for computing generalized Drude weights and quantum relaxation dynamics resulting from $U(1)$-symmetry breaking quantum quenches in the regime of linear response.

\section{The model and representations of \uq}

\subsection{Anisotropic Heisenberg model and quantum group \uq}
Denoting by $\Hp = \CC^2$ the local physical spin-$1/2$ space, the Hamiltonian of the anisotropic Heisenberg spin-$1/2$ chain consisting of $n$ particles can be thought of as an operator in $\End(\Hp^{\otimes n})$,
\begin{align}
H_{XXZ}=\sum_{j=0}^{n-1} h_{j,j+1},
\end{align}
with local interaction terms acting nontrivially on a pair of adjecent sites. Periodic boundary conditions are imposed by taking indices modulo $n$. Operator term indices $j$, $j+1$ denote local physical spaces in the chain, on which the term $h_{j,j+1}$ acts nontrivially. Thus $h_{j,j+1}$ is a trivial extension of the local interaction $h\in \End(\Hp \otimes \Hp)$, which, in our case, takes the form
\begin{align}
h = 2\,\sigma^+ \otimes \sigma^- + 2\,\sigma^-\otimes \sigma^+ + \Delta\,\sigma^\z \otimes \sigma^\z,\label{density}
\end{align}
onto $\End(\Hp^{\otimes n})$. Here, as in standard notation, $\sigma^\alpha$, $\alpha\in\{0,\x,\y,\z\}$ denote Pauli matrices, $\sigma^0=\one$ the identity in $\End(\Hp)$, and additionaly we have $\sigma^\pm \equiv \tfrac{1}{2}(\sigma^\x \pm \ii\sigma^\y)$. In the gapless regime, $|\Delta|\le1$, which will be the subject of our consideration, one can rewrite the parameter of anisotropy as $\Delta = \cos\eta$, introducing a new parameter $\eta\in[0,2\pi)$. Setting $q=e^{i\eta}$, we obtain an intrinsic connection between Heisenberg spin model and the quantum group $\uq$ \cite{fad,pro,pereira,pasquier}. The latter is a Hopf algebra, generated by elements $\mm{S}^+$, $\mm{S}^-$, $q^{2\,\mm{S}^\z}$, satisfying a set of algebraic relations\footnote{See, for example, Ref. \cite{kassel}.}
\begin{align}
\begin{aligned}
&q^{2\,\mm{S}^\z}q^{-2\,\mm{S}^\z}=q^{-2\,\mm{S}^\z}q^{2\,\mm{S}^\z}=1,\\[1em]
&\mm{S}^+\,q^{2\,\mm{S}^\z}=q^{-2}\,q^{2\,\mm{S}^\z}\,\mm{S}^+,\\[1em]
&\mm{S}^-\,q^{2\,\mm{S}^\z}=q^2\,q^{2\,\mm{S}^\z}\,\mm{S}^-,\\[.5em]
&[\mm{S}^+,\mm{S}^-]=\frac{q^{2\,\mm{S}^\z}-q^{-2\,\mm{S}^\z}}{q-q^{-1}}=\frac{\sin(2\,\eta\,\mm{S}^\z)}{\sin\eta}.\label{uq}
\end{aligned}
\end{align}
Here we have used common notation, allowing to write the Lax operator\footnote{Note that in our matrix notation the roles of the auxilliary and the (quantum) physical spaces are interchanged with respect to the literature on quantum inverse scattering method \cite{fad,kbi93}.} as described in \cite{pro},
\begin{align}
\mm{L}(\varphi) = 
\begin{pmatrix}
\sin(\varphi+\eta\,\mm{S}^\z) & (\sin\eta)\,\mm{S}^- \\
(\sin\eta)\,\mm{S}^+ & \sin(\varphi-\eta\,\mm{S}^\z) \\
\end{pmatrix} = \sum_{\mu \in\mathcal{J}} \sigma^{\,\mu}\otimes \mm{L}^{\,\mu}(\varphi)\in \End(\Hp\otimes\Va),\label{lax}
\end{align}
where $\mathcal{J}=\{+,-,0,\z\}$ is the index set and $\Va$ the auxiliary space -- the representation space (module) of the Hopf algebra. The Lax components $\mm{L}^{\,\mu}(\varphi)$ are given by
\begin{align}
\begin{aligned}
&\mm{L}^0(\varphi)=\sin\varphi\,\cos (\eta\mm{S}^\z),\\
&\mm{L}^\z(\varphi)=\cos\varphi\,\sin (\eta\mm{S}^\z),\\
&\mm{L}^\pm(\varphi)=(\sin\eta)\,\mm{S}^\mp.\label{laxcomponents}
\end{aligned}
\end{align}
One can now construct continuously parametrized ($\varphi\in\CC$) \emph{transfer operators} (transfer matrices) $V_n(\varphi)\in\End(\Hp^{\otimes n}),$\footnote{Partial tensor product with respect to physical space, $\otimes_{\rm p}$ is equivalent to trivial expansion of all operators onto $\Hp^{\otimes n}\otimes\Va$ and their subsequent multiplication.}
\begin{align}
V_n(\varphi)=\tr_{\rm a}(\mm{L}(\varphi)^{\otimes_{\rm p} n}).\label{transfer}
\end{align}
In general their conservation and involution follows from the quantum Yang-Baxter equation, evaluated in a representation over a triple tensor product $\Hp\otimes\Hp\otimes\Va$ and $\Hp\otimes\Va\otimes\Va$, respectively \cite{fad,pro}. First let us describe the explicit form of the representations used.

\subsection{Irreducible representations of \uq}
The theory of irreducible representations of \uq allows for an especially rich family of representation structures at the roots of unity, where enlarged center of the algebra provides additional representation parameters \cite{kac}. Here, besides the highest-weight representations, more exotic structures, e.g. (semi)cyclic representations, are encountered. For $q$ taking values from the set of roots of unity, the anisotropy parameter $\Delta$ densely populates the interval $[-1,1]$. Let us denote by $d$ the order of root of unity, the lowest nontrivial natural number such that $q^d=1$. Setting
\begin{align}
m=
\begin{cases}
d,\;d\;\text{odd}\\
\frac{d}{2},\;d\;\text{even} 
\end{cases},\label{dimension}
\end{align}
one finds that $m$ is the highest possible dimension of irreducible representation of \uq at $q$ - root of unity \cite{kassel}. We consider the following irreducible representations, preserving the form of the Lax operator \eqref{lax} and up to isomoprhism equivalent to the classified ones \cite{kassel, arnaudon}:
\small
\begin{align}
\Va_1&(s,\alpha_1,\beta_1),\;\;s,\alpha_1,\beta_1\in\CC\notag\\
&\begin{aligned}
&\mm{S}^\z_1=\sum_{k=0}^{m-1}(s-k)\ket{k}\bra{k},\\
&\mm{S}^+_1=\sum_{k=0}^{m-2}\left(\frac{\sin(k+1)\eta}{\sin\eta}+\frac{\alpha_1\,\beta_1\sin\eta}{\sin(2s-k)\eta}\right)\ket{k}\bra{k+1}+\alpha_1\ket{m-1}\bra{0},\\
&\mm{S}^-_1=\sum_{k=0}^{m-2}\frac{\sin(2s-k)\eta}{\sin\eta}\ket{k+1}\bra{k}+\beta_1\ket{0}\bra{m-1},\label{cycl1}
\end{aligned}\\[1em]
\Va_2&(s,\alpha_2,\beta_2),\;\;s,\alpha_2,\beta_2\in\CC\notag\\
&\begin{aligned}
&\mm{S}^\z_2=\sum_{k=0}^{m-1}(s-k)\ket{k}\bra{k},\\
&\mm{S}^+_2=\sum_{k=0}^{m-2}\left(\frac{\sin(2s-k)\eta}{\sin\eta}+\frac{\alpha_2\,\beta_2\sin\eta}{\sin(k+1)\eta}\right)\ket{k}\bra{k+1}+\alpha_2\ket{m-1}\bra{0},\\
&\mm{S}^-_2=\sum_{k=0}^{m-2}\frac{\sin(k+1)\eta}{\sin\eta}\ket{k+1}\bra{k}+\beta_2\ket{0}\bra{m-1},\label{cycl2}
\end{aligned}\\[1em]
\Va_3&(p,\gamma),\;\;\gamma\in\CC,\;\;p\in\ZZ,\;\; 0\le p \le m-2\notag\\
&\begin{aligned}
&\mm{S}^\z_3=\sum_{k=0}^{m-2}\left(k-\frac{p}{2}\right)\ket{k}\bra{k}-(\frac{p}{2}+1)\ket{m-1}\bra{m-1},\\
&\mm{S}^-_3=\sum_{k=0}^{m-2}\frac{\sin(p-k)\eta}{\sin\eta}\ket{k}\bra{k+1},\\
&\mm{S}^+_3=\sum_{k=0}^{m-2}\frac{\sin(k+1)\eta}{\sin\eta}\ket{k+1}\bra{k}+\gamma\ket{0}\bra{m-1}.\label{cycl3}
\end{aligned}
\end{align}
\normalsize
The first two of these structures are isomorphic if we set
\begin{align}
\begin{aligned}
\beta_2=\left(\prod_{l=1}^{m-1}\frac{\sin(2s-l+1)\eta}{\sin l\eta}\right)\,\beta_1,\\
\alpha_1=\left(\prod_{l=1}^{m-1}\frac{\sin(2s-l+1)\eta}{\sin l\eta}\right)\,\alpha_2.
\end{aligned}
\end{align}
The \uq-module isomorphism is given by the adjoint action of $\mm{U}: \Va_2\rightarrow \Va_1$,
\begin{align}
\mm{U} =\ket{0}\bra{0} + \sum_{k=1}^{m-1}\prod_{l=1}^k\frac{\sin(2s-l+1)\eta}{\sin l\eta}\,\ket{k}\bra{k}.\label{isomorphism}
\end{align}
Indeed, one can easily check that for all $\alpha\in\{+,-,\z\}$, relation 
\begin{align}
\mm{S}^\alpha_1=\mm{U}\,\mm{S}^\alpha_2\,\mm{U}^{-1}
\end{align}
holds. Note, however, that $\mm{U}$ does not define an isomorphism for all possible values of $s$. A counterexample, which will be relevant in our case, is $s=0$. The highest-weight representation $\Va_1(s,0,0)$ was used in Ref.~\cite{pro}. In what follows, we will leave out the representation indices in operator definitions ($\mm{S}^\alpha_j=\mm{S}^\alpha$). Parameters $\alpha,\,\beta,\,\gamma$, couple the highest-weight vector with the lowest-weight one. Therefore, they shall henceforth be referred to as \emph{coupling parameters}. If one of $\alpha,\beta$ is non-zero, the representation is called \emph{semicyclic}. If both are nonzero, the representation is refered to as \emph{cyclic}. The representation $\Va_3(p,\gamma\ne 0)$ is \emph{a priori} semicyclic. Coupling of the highest and the lowest-weight vectors leads to  a problem with the \emph{fundamental commutation relation} used for a proof of conservation of transfer operators \eqref{transfer}. We discuss this problem and its consequences in the next subsection.

\subsection{The problem of the fundamental commutation relation}
Conservation of transfer operators \eqref{transfer} follows from the train argument \cite{fad} used on the fundamental commutation relation (FCR), i.e. the Yang-Baxter equation evaluated on the triple tensor product $\Hp\otimes\Hp\otimes\Va$,
\begin{align}
\mm{R}_{12}(x/y)\mm{L}_{13}(x)\mm{L}_{23}(y)=\mm{L}_{23}(y)\mm{L}_{13}(x)\mm{R}_{12}(x/y)\label{fcr}.
\end{align}
Here $x,\, y\in\CC\setminus\{0\}$ are the spectral parameters, in our case taking an explicit form $x=e^{i\varphi}$, $y=e^{i\varphi'}$ and operator indices\footnote{For example $\mm{R}_{12}=\mm{R}\otimes 1$, where $\mm{R}\in\End(\Hp\otimes\Hp)$ is given by \eqref{rmatrix}.} refer to vector spaces in the triple tensor product, on which the operators act nontrivially \cite{fad}.\footnote{We note that all representation modules discussed here are to be understood in the sense of so-called {\em evaluation map representation} of the infinite-dimensional loop algebra $\uLq$.} The form of Lax operators $\mm{L}$ is given by \eqref{lax} and the \emph{trigonometric R-matrix} by
\begin{align}
\mm{R}(x)=
\begin{pmatrix}
qx-q^{-1}x^{-1} & 0 & 0 & 0\\
0 & x-x^{-1} & q-q^{-1} & 0\\
0 & q-q^{-1} & x-x^{-1} & 0\\
0 & 0 & 0 & qx-q^{-1}x^{-1}
\end{pmatrix}\in \End(\Hp\otimes\Hp).\label{rmatrix}
\end{align}

One can prove that FCR \eqref{fcr} holds for $R$-matrix \eqref{rmatrix} and Lax operator \eqref{lax} iff the following set of relations is satisfied \cite{fad}: 
\begin{align}
\begin{aligned}
&\mm{S}^+\,q^{\mm{S}^\z}=q^{-1}\,q^{\mm{S}^\z}\,\mm{S}^+,\\[1em]
&\mm{S}^-\,q^{\mm{S}^\z}=q\,q^{\mm{S}^\z}\,\mm{S}^-,\\[0.5em]
&[\mm{S}^+,\mm{S}^-]=\frac{q^{2\,\mm{S}^\z}-q^{-2\,\mm{S}^\z}}{q-q^{-1}}=\frac{\sin(2\,\eta\,\mm{S}^\z)}{\sin\eta}.\label{nuq}
\end{aligned}
\end{align}
Notice the essential difference between \uq-defining relations \eqref{uq} and the Eqs.~\eqref{nuq}. A question that now arises is, in which case are the latter relations satisfied if the former hold. This question has been discussed in relation to the existence of intertwiner for representations $\Hp$ and $\Va_j$, $j\in\{1,2,3\}$, in Ref.~\cite{korff}. In general we can summarize the action of \uq-generators as
\begin{align}
\begin{aligned}
&q^{\mm{S}^\z}\ket{k}=f^\z(k)\,\ket{k},\;\;k=0,1,2,3,...,m-1,\\[.5em]
&\mm{S}^+\ket{k+1}=f^+(k)\,\ket{k},\;\;k=0,1,2,3,...,m-2,\\[.5em]
&\mm{S}^-\ket{k}=f^-(k)\,\ket{k+1},\;\;k=0,1,2,3,...,m-2,\\[.5em]
&\mm{S}^+\ket{0}=\alpha\,\ket{m-1},\\[.5em]
&\mm{S}^-\ket{m-1}=\beta\,\ket{0},\label{ansatz}
\end{aligned}
\end{align}
for some discrete maps $f^+,\,f^-,\,f^\z: I\rightarrow \CC$, with $f^+(m-1)=\alpha$, $f^-(m-1)=\beta$, defined on the index set $I=\{0,1,2,...,m-1\}$. The relations \eqref{nuq} and \eqref{ansatz} result in the following set of equations:
\begin{align}
\begin{aligned}
&(\mm{S}^+q^{\mm{S}^\z}-q^{-1}q^{\mm{S}^\z} \mm{S}^+)\,\ket{k}=(f^\z(k)-q^{-1}f^\z(k-1))\;\mm{S}^+\ket{k}=0,\;k=1,2,...,m-1,\\[.5em]
&(\mm{S}^+q^{\mm{S}^\z}-q^{-1}q^{\mm{S}^\z} \mm{S}^+)\,\ket{0}=(f^\z(0)-q^{-1}f^\z(m-1))\;\mm{S}^+\ket{0}=0,\\[.5em]
&(\mm{S}^-q^{\mm{S}^\z}-q\,q^{\mm{S}^\z}\,\mm{S}^-)\,\ket{k}=(f^\z(k)-q\,f^\z(k+1))\,\mm{S}^-\;\ket{k}=0,\;k=0,1,...,m-2,\\[.5em]
&(\mm{S}^-q^{\mm{S}^\z}-q\,q^{\mm{S}^\z}\,\mm{S}^-)\,\ket{m-1}=(f^\z(m-1)-q\,f^\z(0))\;\mm{S}^-\ket{m-1}=0.
\end{aligned}
\end{align}
A recursion relation then follows for the index map $f^\z$,
\begin{align}
\begin{aligned}
&f^\z(k+1)=q^{-1}f^\z(k),\\
&f^\z(m-1)=q\,f^\z(0),
\end{aligned}
\end{align}
implying $q^m=1$, since $f^\z$ is nonzero. Note that in case of a representation with both the highest and the lowest-weight vector, there is no such constraint on the module dimension. Using Eq.~\eqref{dimension}, one can conclude, that the dimension of semicyclic or cyclic irreducible representations of \uq, for which FCR~\eqref{fcr} holds, is always odd.

If one considers FCR as the sole source of conservation of transfer operators \eqref{transfer},
then the construction of conservation laws from the semicyclic irreducible representations is invalid for $q$ being a root of unity of even order. This restricts the values of the anisotropy parameter $\Delta \in [-1,1]$ one can discuss using these representations, onto the following dense countable set
\begin{align}
S=\left\{\cos\left(\frac{2\,l}{2\,k-1}\,\pi\right)\right\}_{k,l\in\mathbbm{N},\;l<k}.
\end{align}
Indeed, explicit symbolic evaluation of commutators $[V_n(\varphi),H_{XXZ}]$ for small systems in case of $\Delta\not\in S$, $\Delta = \cos\eta$, where $q = e^{i\eta}$ is a root of unity, yields non-zero results. 
Note that although this set densely covers the interval $[-1,1]$, it does not so in a symmetric fashion with respect to $\Delta =0$. For example, parameter $\Delta=-1/2$ ($\eta=2\pi/3$) is included, while $\Delta=1/2$ ($\eta=\pi/3$) is not. This can be amended in the thermodynamic limit by noting, that the adjoint action of unitary operator $U\in \End(\Hp^{\otimes n})$, $n\in 2\mathbbm{N}$ given by
\begin{align}
U=(\one\otimes\sigma^\z)^{\otimes\, n/2},
\end{align}
is equivalent to flipping the sign of the anisotropy parameter $\Delta$, i.e. $UH_{XXZ}(\Delta)U=-H_{XXZ}(-\Delta)$. Hence we have an equivalence
\begin{align}
[H_{XXZ}(\Delta),Y] = 0\;\;\Leftrightarrow\;\;[H_{XXZ}(-\Delta), UYU]=0,
\end{align}
i.e. if $Y$ is a conserved charge in case of anisotropy $\Delta$, then $\tilde{Y}=UYU$ is a conserved charge for anisotropy $-\Delta$.

\subsection{About involution}

For $q$ equal to a root of unity there is, in general, no solution to the Yang-Baxter equation in the algebraic sense \cite{arnaudon}. From Arnaudon's discussion of validity of Yang-Baxter equation \cite{arnaudon} and the existence of intertwiner at $s'=s=0$ \cite{korff}, we can conclude that 
\begin{align}
[V_n(\varphi)\bigr|_{s=0},V_n(\varphi')\bigr|_{s'=0}]=0
\end{align}
holds in case $q^m=1$, $m=\text{odd}$, where $m$ is the dimension of irreducible representation. In case $q^m=-1$ the symbolic computation for small system sizes yields non-vanishing commutators among transfer matrices.

\section{Quasilocal conserved quantities}
\label{qcq}

\subsection{Construction}
In this section we explicitly describe the construction of new conservation laws, exhibiting an additional property called \emph{quasilocality}. For completeness, let us revisit the definition of quasilocality, as stated in Ref.~\cite{pro}.
\newdefinition{defin}{Definition}
\begin{defin}
An operator sequence $\{Y_n\}_{n\in\mathbbm{N}}$, $Y_n\in \End(\Hp^{\otimes n})$ which can be written as 
\begin{align}
Y_n=\sum_{r\le n}\sum_{x=0}^{n-1}\mathcal{S}^x(q_r\otimes\one^{\otimes (n-r)}),
\end{align}
with operator terms $q_r\in \End(\Hp^{\otimes r})$ satisfying
\begin{align}
||q_r||_{\scriptscriptstyle HS}\le \gamma e^{-\zeta r},\label{qquasi}
\end{align}
for some positive constants $\gamma, \zeta > 0$, is called \emph{quasilocal}.\footnote{By $||-||_{\scriptscriptstyle HS}$, we have here denoted a vector norm of an operator, naturally defined by the \emph{Hilbert-Schmidt} scalar product,
$\left(-,-\right): \End(\Hp^{\otimes k})\rightarrow \CC$, $\left(A,B\right)=\tr(A^\dagger B)/2^k$.}
\end{defin}

As has been seen in the preceding section, for $\Delta\in S$, the conservation and commutation of transfer operators \eqref{transfer} is guaranteed. Transfer operators inherit a dependence on parameters of the representation from the Lax operators they consist of. As was implicitly hinted above, the representation parameters of our interest are the ones, coupling the highest and the lowest-weight vectors, thus making the representation (semi)cyclic. The task is now to make transfer operators quasilocal. In order to achieve this, we will differentiate the transfer operator with respect to the complex coupling parameter $\beta$ (or $\alpha$ or $\gamma$) and then set all the representation parameters, i.e. all parameters except $\varphi$: the coupling parameter $\beta$ (or $\alpha$ or $\gamma$) and the \emph{spin} parameter $s$ (or $p$), to zero. The representations $\Va_1$ and $\Va_2$ are related by the map given by \eqref{isomorphism}. Since for our choice of representation parameters ($s = 0$), this map is no longer an isomorphism, we do not expect exact equivalence of the resulting conservation laws.

Let us define rescaled Lax components as $\mm{\tilde L}^{\,\mu}_0(\varphi)=\csc\varphi\,\mm{L}^{\,\mu}(\varphi)\bigr|_{s=0}$, where $\mm{L}^{\,\mu}(\varphi)$ are given by Eqs.~\eqref{laxcomponents} and denote $s_k=\sin(k\eta)$, $c_k = \cos(k\eta)$. Note that $\mm{L}^{\,\mu}(\varphi)$ posesses an implicit dependence on both the spin and the coupling parameter and $\mm{\tilde L}^{\,\mu}_0(\varphi)$, on the other hand retains only implicit dependence on the coupling parameter. As will soon become clear, this procedure gives nontrivial results only in the following cases:
\newcommand\NoIndent[1]{%
  \par\vbox{\parbox[t]{\linewidth}{#1}}%
}
\begin{itemize}
\item[--] $\Va_1(s,\alpha,0)$, differentiation with respect to $\alpha$:
\newline
\NoIndent{
\begin{minipage}[b]{0.15\linewidth}
     \raisebox{0.2\height}{\includegraphics[width=50pt]{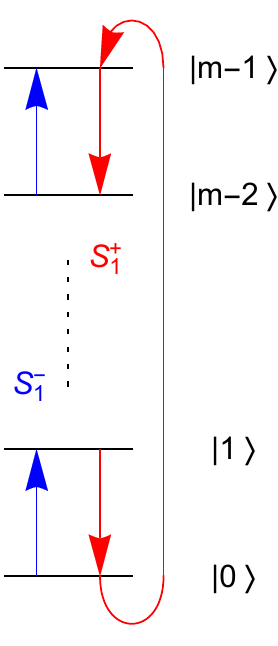}}
\end{minipage}
\hspace{.5cm}
\begin{minipage}[b]{0.875\linewidth}
\begin{align}
\begin{aligned}
&\mm{\tilde L}^0_0(\varphi)\bigr|_{\alpha=0}=\sum_{k=0}^{m-1}c_k\ket{k}\bra{k},\\[.25em]
&\mm{\tilde L}^\z_0(\varphi)\bigr|_{\alpha=0}=-\cot\varphi\sum_{k=0}^{m-1}s_k\ket{k}\bra{k},\\[.25em]
&\mm{\tilde L}^+_0(\varphi)\bigr|_{\alpha=0}=-\csc\varphi\sum_{k=0}^{m-2}s_k\ket{k+1}\bra{k},\\[.25em]
&\mm{\tilde L}^-_0(\varphi)\bigr|_{\alpha=0}=\csc\varphi\sum_{k=0}^{m-2}s_{k+1}\ket{k}\bra{k+1},\\[.75em]
&\partial_\alpha\mm{\tilde L}^-_0(\varphi)\bigr|_{\alpha=0}=\sin\eta\,\ket{m-1}\bra{0},
\label{lax1}
\end{aligned}
\end{align}
\end{minipage}
}

\item[--] $\Va_2(s,0,\beta)$, differentiation with respect to $\beta$:
\newline
\NoIndent{
\begin{minipage}[b]{0.15\linewidth}
    \raisebox{0.3\height}{\includegraphics[width=50pt]{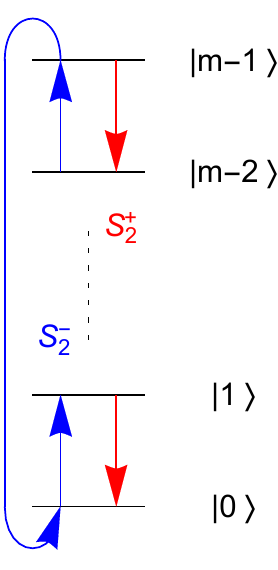}}
\end{minipage}
\hspace{.5cm}
\begin{minipage}[b]{0.875\linewidth}
\begin{align}
\begin{aligned}
&\mm{\tilde L}^0_0(\varphi)\bigr|_{\beta=0}=\sum_{k=0}^{m-1}c_k\ket{k}\bra{k},\\[.25em]
&\mm{\tilde L}^\z_0(\varphi)\bigr|_{\beta=0}=-\cot\varphi\sum_{k=0}^{m-1}s_k\ket{k}\bra{k},\\[.25em]
&\mm{\tilde L}^+_0(\varphi)\bigr|_{\beta=0}=\csc\varphi\sum_{k=0}^{m-2}s_{k+1}\ket{k+1}\bra{k},\\[.25em]
&\mm{\tilde L}^-_0(\varphi)\bigr|_{\beta=0}=-\csc\varphi\sum_{k=0}^{m-2}s_k\ket{k}\bra{k+1},\\[.75em]
&\partial_\beta\mm{\tilde L}^+_0(\varphi)\bigr|_{\beta=0}=\sin\eta\,\ket{0}\bra{m-1},
\label{lax2}
\end{aligned}
\end{align}
\end{minipage}
}

\item[--] $\Va_3(p,\gamma)$, differentiation with respect to $\gamma$:
\newline
\NoIndent{
\begin{minipage}[b]{0.15\linewidth}
    \raisebox{0.3\height}{\includegraphics[width=50pt]{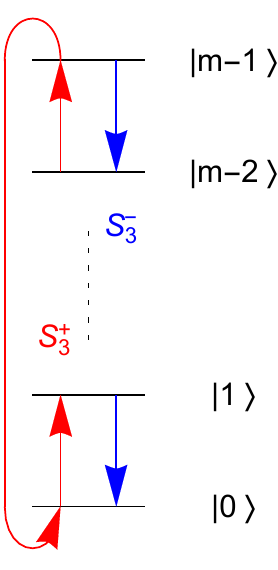}}
\end{minipage}
\hspace{.5cm}
\begin{minipage}[b]{0.875\linewidth}
\begin{align}
\begin{aligned}
&\mm{\tilde L}^0_0(\varphi)\bigr|_{\gamma=0}=\sum_{k=0}^{m-1}c_k\ket{k}\bra{k},\\[.25em]
&\mm{\tilde L}^\z_0(\varphi)\bigr|_{\gamma=0}=\cot\varphi\sum_{k=0}^{m-1}s_k\ket{k}\bra{k},\\[.25em]
&\mm{\tilde L}^+_0(\varphi)\bigr|_{\gamma=0}=-\csc\varphi\sum_{k=0}^{m-2}s_k\ket{k}\bra{k+1},\\[.25em]
&\mm{\tilde L}^-_0(\varphi)\bigr|_{\gamma=0}=\csc\varphi\sum_{k=0}^{m-2}s_{k+1}\ket{k+1}\bra{k},\\[.75em]
&\partial_\gamma\mm{\tilde L}^-_0(\varphi)\bigr|_{\gamma=0}=\sin\eta\,\ket{0}\bra{m-1}.
\label{lax3}
\end{aligned}
\end{align}
\end{minipage}
}
\end{itemize}

For the sake of simplicity let us denote
\begin{align}
\label{dodatno}
\mm{L}^{\,\mu}_0(\varphi)=\mm{\tilde L}^{\,\mu}_0(\varphi)\bigr|_{\zeta=0},
\hspace{2cm}
\partial_\zeta\mm{L}^{\,\mu}_0(\varphi)=\partial_\zeta\mm{\tilde L}^{\,\mu}_0(\varphi)\bigr|_{\zeta=0},
\end{align}
where $\zeta$ stands for $\alpha$, $\beta$ or $\gamma$. $\mm{L}^{\,\mu}_0(\varphi)$ now depends only on the explicit complex (spectral) parameter $\varphi$. 
\newdefinition{exm}{Example}
\begin{exm}
As an example of construction of quasilocal operators, we take a look at the representation  $\Va_2(s,0,\beta)$. Using the \emph{periodic left-shift} operator 
\begin{align}
\hat{\mathcal{S}}(\sigma^{\alpha_0}\otimes \sigma^{\alpha_1} \otimes\cdots\sigma^{\alpha_{n-2}}\otimes\sigma^{\alpha_{n-1}}) &= \sigma^{\alpha_{1}}\otimes \sigma^{\alpha_2}\otimes\cdots\sigma^{\alpha_{n-1}}\otimes\sigma^{\alpha_{0}},
\end{align}
we can introduce quasilocal conserved charges $X_n(\varphi)$
\begin{align}
X_n(\varphi)&=\frac{\partial_\beta V_n(\varphi)\bigr|_{s=0,\,\beta=0}}{(\sin\eta)^2\,(\sin\varphi)^{n-2}}=\left(\frac{\sin\varphi}{\sin\eta}\right)^2\sum_{\mu_1,...,\mu_n\in\mathcal{J}}\tr{\left(\partial_\beta\mm{L}_0^{\,\mu_1}(\varphi)\,\mm{L}_0^{\,\mu_2}(\varphi)\cdots\mm{L}_0^{\,\mu_n}(\varphi)\right)}\,\times\notag\\
&\times\sum_{x=1}^n \hat{\mathcal{S}}^x(\sigma^{\,\mu_1}\otimes\sigma^{\,\mu_2}\otimes\cdots\sigma^{\,\mu_{n}}),\label{monodromyderivative}
\end{align}
using Eqs.~\eqref{lax2}, \eqref{dodatno}. Defining local operator terms 
\begin{align}
&q^{\scriptscriptstyle (2)}_r(\varphi)=\sum_{\mu_2,...\mu_{r-1}\in\mathcal{J}}\bra{m-1}\mm{L}_0^{\,\mu_2}(\varphi)\cdots\mm{L}_0^{\,\mu_{r-1}}(\varphi)\ket{1}\sigma^+\otimes\sigma^{\,\mu_2}\otimes\cdots\sigma^{\,\mu_{r-1}}\otimes\sigma^+,\label{q2}
\end{align}
for $r\ge2$, we can rewrite equation \eqref{monodromyderivative} as
\begin{align}
X_n(\varphi)=\sum_{r=2}^n\sum_{x=0}^{n-1}\hat{\mathcal{S}}^x(q^{\scriptscriptstyle (2)}_r(\varphi)\otimes\mathbbm{1}^{\otimes (n-r)}).\label{family2}
\end{align}
The superscript index $(2)$ labels the representation $\Va_2(s,0,\beta)$.
Note from \eqref{lax2} and \eqref{q2}, that operator terms $q^{\scriptscriptstyle (2)}_r(\varphi)$ do not conserve $z$-component of magnetization, since the sequence of Lax operators in each coefficient needs to couple the \emph{ket} vector $\ket{1}$ with the \emph{bra} vector $\bra{m-1}$ and thus requires a surplus of $m-2$ operators $\mm{L}_0^+(\varphi)$. Consequentially, only the terms with $r\ge m$ remain in (\ref{family2}). Taking into account the two boundary $\sigma^+$ matrices, each term of $q^{\scriptscriptstyle (2)}_r(\varphi)$ thus consists of a surplus of $m$  matrices $\sigma^+$. 

Quasilocality of operator sequence $\{X_n(\varphi)\}_n$ follows from the argument given in section 5 of Ref. \cite{pro}, for spectral parameter $\varphi$ satisfying the constraint $|\Re \varphi-\tfrac{\pi}{2}|<\tfrac{\pi}{2m}$. The argument is based on the estimation of the spectral radius of an auxilliary transfer matrix $\mm{T(\varphi,\varphi')}\in{\rm lsp}\{\ket{k};k=1,\ldots,m-1\}$, given by 
\begin{align}
 \mm{T}(\varphi,\varphi') &=&\!\!\!\sum_{k=1}^{m-1} (c^2_k +\cot\varphi\cot\varphi' s^2_k)\ket{k}\!\bra{k} + \sum_{k=1}^{m-2}\frac{|s_{k}s_{k+1}|}{2\sin\varphi\sin\varphi'}\left(\ket{k}\!\bra{k\!+\!1} + \ket{k\!+\!1}\!\bra{k}\right),\label{tt1}
\end{align}
satisfying
\begin{align}
\frac{1}{2^{r}}\tr\left( [q^{\scriptscriptstyle (2)}_r(\varphi)]^T q^{\scriptscriptstyle (2)}_r(\varphi')\right) = \frac{1}{4}\bra{m-1} \mm{T}(\varphi,\varphi')^{r-2}\ket{1}, \quad { r \ge m}.
\label{tt2}
\end{align}
Here we should point out the difference in the bra vector in \eqref{tt2} and equation (53) of the Ref. \cite{pro}.\footnote{One needs to slightly alter the proof for each particular choice of representation. When evaluating the LHS of Eq.~\eqref{tt2}, the correct correspondence between ${\rm lsp}\{\ket{k}\otimes\ket{k};k=1,\ldots,m-1\}$ and ${\rm lsp}\{\ket{k};k=1,\ldots,m-1\}$ must be chosen, in order to arrive at the transfer matrix of the form \eqref{tt1}. See Ref. \cite{pro} for comparison.} Since the local operator terms $q_r^{\scriptscriptstyle (2)}(\varphi)$ are Hilbert-Schmidt orthogonal with respect to $r$, the following estimation can be produced:
\begin{align}
\left(X_n(\bar{\varphi}),X_n(\varphi')\right) \le n\sum_{r=2}^\infty 
\frac{1}{2^{r}}\tr\left( [q^{(2)}_r(\varphi)]^T q^{(2)}_r(\varphi')\right) 
=n\,K(\varphi,\varphi'),
\end{align}
with
\begin{align}
K(\varphi,\varphi')=\frac{1}{4}\bra{m-1}(\one-\mm{T}(\varphi,\varphi'))^{-1}\ket{1}=\frac{1}{4}\,\psi_{m-1},
\end{align}
where $\ket{\psi}$ solves the equation $(\one-\mm{T}(\varphi,\varphi'))\ket{\psi}=\ket{1}$. Explicit calculation of all components $\psi_j$, $j\in\{1,\cdots, m-1\}$ has been done in Ref. \cite{pro}. The final result for the so-called Hilbert-Schmidt kernel in our case reads:
\begin{align}
K(\varphi,\varphi')=\lim_{n\rightarrow\infty}\tfrac{1}{n}\left(X_n(\bar\varphi),X_n(\varphi')\right)=\frac{\sin\varphi\,\sin\varphi'\,\sin(\varphi+\varphi')}{2\,(\sin\eta)^2\,\sin(m(\varphi+\varphi'))}\label{hsnorm}
\end{align}

As an explicit example let us write down full symbolic expressions of the first few local operator terms $q^{\scriptscriptstyle (2)}_3(\varphi)$, $q^{\scriptscriptstyle (2)}_4(\varphi)$ and $q^{\scriptscriptstyle (2)}_5(\varphi)$ for $\eta=2\pi/3$ ($\Delta=-1/2$):
\footnotesize
\begin{align}
\begin{aligned}
q^{\scriptscriptstyle (2)}_3(\varphi)&=-\frac{\sqrt{3}}{2}\csc\varphi\,\left(\sigma^+\otimes\sigma^+\otimes\sigma^+\right),\\[1em]
q^{\scriptscriptstyle (2)}_4(\varphi)&=\frac{\sqrt{3}}{4}\csc\varphi\,\left(\sigma^+\otimes\mathbbm{1}\otimes\sigma^+\otimes\sigma^+\right)+\frac{\sqrt{3}}{4}\csc\varphi\,\left(\sigma^+\otimes\sigma^+\otimes\mathbbm{1}\otimes\sigma^+\right)+\\[1em]
&+\frac{3}{4}\cot\varphi\,\csc\varphi\,\left(\sigma^+\otimes\sigma^+\otimes\sigma^\z\otimes\sigma^+\right)
-\frac{3}{4}\cot\varphi\,\csc\varphi\,\left(\sigma^+\otimes\sigma^\z\otimes\sigma^+\otimes\sigma^+\right),\\[1em]
q^{\scriptscriptstyle (2)}_5(\varphi)&=-\frac{\sqrt{3}}{8}\csc\varphi\,\left(\sigma^+\otimes\one\otimes\one\otimes\sigma^+\otimes\sigma^+\right)-\frac{\sqrt{3}}{8}\csc\varphi\,\left(\sigma^+\otimes\one\otimes\sigma^+\otimes\one\otimes\sigma^+\right)-\\[1em]
&-\frac{\sqrt{3}}{8}\csc\varphi\,\left(\sigma^+\otimes\sigma^+\otimes\one\otimes\one\otimes\sigma^+\right)-
\frac{3\sqrt{3}}{8}(\csc\varphi)^3\,\left(\sigma^+\otimes\sigma^+\otimes\sigma^-\otimes\sigma^+\otimes\sigma^+\right)-\\[1em]
&-\frac{3}{8}\cot\varphi\,\csc\varphi\,\left(\sigma^+\otimes\one\otimes\sigma^+\otimes\sigma^\z\otimes\sigma^+\right)+
\frac{3}{8}\cot\varphi\,\csc\varphi\,\left(\sigma^+\otimes\one\otimes\sigma^\z\otimes\sigma^+\otimes\sigma^+\right)-\\[1em]
&-\frac{3}{8}\cot\varphi\,\csc\varphi\,\left(\sigma^+\otimes\sigma^+\otimes\one\otimes\sigma^\z\otimes\sigma^+\right)
-\frac{3}{8}\cot\varphi\,\csc\varphi\,\left(\sigma^+\otimes\sigma^+\otimes\sigma^\z\otimes\one\otimes\sigma^+\right)+\\[1em]
&+\frac{3}{8}\cot\varphi\,\csc\varphi\,\left(\sigma^+\otimes\sigma^\z\otimes\one\otimes\sigma^+\otimes\sigma^+\right)
+\frac{3}{8}\cot\varphi\,\csc\varphi\,\left(\sigma^+\otimes\sigma^\z\otimes\sigma^+\otimes\one\otimes\sigma^+\right)-\\[1em]
&-\frac{3\sqrt{3}}{8}(\cot\varphi)^2\,\csc\varphi\,\left(\sigma^+\otimes\sigma^+\otimes\sigma^\z\otimes\sigma^\z\otimes\sigma^+\right)+\\[1em]
&+\frac{3\sqrt{3}}{8}(\cot\varphi)^2\,\csc\varphi\,\left(\sigma^+\otimes\sigma^\z\otimes\sigma^+\otimes\sigma^\z\otimes\sigma^+\right)-\\[1em]
&-\frac{3\sqrt{3}}{8}(\cot\varphi)^2\,\csc\varphi\,\left(\sigma^+\otimes\sigma^\z\otimes\sigma^\z\otimes\sigma^+\otimes\sigma^+\right).\label{example}
\end{aligned}
\end{align}
\normalsize
\end{exm}

\begin{exm}
Since in case of representation $\Va_1(s,0,\beta)$ the Lax component $\mm{\tilde L}_0^+(\varphi)$ destroys the ket vector $\ket{0}$, one cannot get nontrivial result by differentiation with respect to $\beta$. However one can take $\Va_1(s,\alpha,0)$ and differentiate with respect to $\alpha$. In this case the operator terms $q^{\scriptscriptstyle (1)}_r(\varphi)$ take the form
\begin{align}
q^{\scriptscriptstyle (1)}_r(\varphi)=\sum_{\mu_2,...\mu_{r-1}\in\mathcal{J}}\bra{1}\mm{L}_0^{\,\mu_2}(\varphi)\cdots\mm{L}_0^{\,\mu_{r-1}}(\varphi)\ket{m-1}\sigma^-\otimes\sigma^{\,\mu_2}\otimes\cdots\sigma^{\,\mu_{r-1}}\otimes\sigma^-,\label{q1}
\end{align}
where Lax components are given by \eqref{lax1}. Here, each coefficient contains a surplus of $m-2$ operators $\mm{L}_0^-(\varphi)$. Defining the spin flip or \emph{parity} operator\footnote{Here, the spin flip will always be denoted by $P$, regardless of the physical space it acts on. Its action will be clear from the context.} $P\in \End(\Hp^{\otimes r})$ as
$P=(\sigma^x)^{\otimes r}$, $P^2 = \one$, we can see, after some rumination, that
\begin{align}
q^{\scriptscriptstyle (1)}_r(\varphi)=Pq^{\scriptscriptstyle (2)}_r(\varphi)P=\left(q^{\scriptscriptstyle (2)}_r(\pi-\varphi)\right)^T.
\end{align}
Therefore the Hilbert-Schmidt norm of such conserved quantities remains the same as in Eq.~\eqref{hsnorm}. 
\end{exm}

\begin{exm}
Taking the third representation with Lax components given by \eqref{lax3} and differentiating the corresponding transfer operator with respect to $\gamma$, we get
\begin{align}
q^{\scriptscriptstyle (3)}_r(\varphi)=\sum_{\mu_2,...\mu_{r-1}\in\mathcal{J}}\bra{m-1}\mm{L}_0^{\,\mu_2}(\varphi)\cdots\mm{L}_0^{\,\mu_{r-1}}(\varphi)\ket{1}\sigma^-\otimes\sigma^{\,\mu_2}\otimes\cdots\sigma^{\,\mu_{r-1}}\otimes\sigma^-,\label{q3}
\end{align}
for operator terms. It is straightforward to see that again 
\begin{align}
q^{\scriptscriptstyle (3)}_r(\varphi)=Pq^{\scriptscriptstyle (2)}_r(\varphi)P=\left(q^{\scriptscriptstyle (2)}_r(\pi-\varphi)\right)^T \equiv q^{\scriptscriptstyle (1)}_r(\varphi)
\end{align}
holds. For an example of explicit form of operator terms, see Eqs. \eqref{example}. 

Therefore, we have -- up to a trivial spin flip -- a single additional family of semi-cyclic quasilocal charges which, remarkably, do not conserve the $z$-magnetization,
\begin{align}
[X_n(\varphi),M^\z_n] \neq 0,\quad M^\z_n = \sum_{j=1}^n \sigma^\z_j.
\end{align}
\end{exm}

\section{Applications}

Let us briefly discuss the possibility of applications. Conserved quantities are especially important for estimation of dynamical susceptibilities of extensive observables, by means of the \emph{Mazur-Suzuki} inequality \cite{pro, proenej, enej, suzuki}. Let $A$ be such an observable with a fixed parity $\nu$, i.e. $PAP=(-1)^\nu A$. As was done in Ref. \cite{pro} for the case of magnetization (or spin) current, the dynamical susceptibility can be bounded from below in accordance with the inequality
\begin{align}
D_A=\lim_{n\rightarrow \infty}\frac{1}{2n}\expect{\overline A}^2\ge\frac{1}{2}\int \d{^2\varphi}\,a(\varphi)f(\varphi).
\label{eq:bound}
\end{align} 
Here $\overline{A}$ denotes a formal time average,
\begin{align}
a(\varphi)=\lim_{n\rightarrow\infty}\frac{1}{n}\left(A,X_n(\varphi)\right)\label{projection}
\end{align}
is a projection of an observable onto the conserved quantity, and $f$ is a complex-valued holomorphic function, defined on $\mathcal{D}_m=\{\varphi\in\CC\,;\; |\Re \varphi-\tfrac{\pi}{2}|<\tfrac{\pi}{2m}\}$, satisfying the Fredholm equation of the first kind,
\begin{align}
\frac{1}{2}\int_{\mathcal{D}_m} \d{^2\varphi'}K(\varphi,\varphi')f(\varphi')=\overline{a(\bar{\varphi})}.\label{fredholmeq}
\end{align}

Since $z$-magnetization $M^\z_n$ is not conserved by our conservation laws, the observables eligible for such a consideration must not conserve it as well in order for $a(\varphi)$ to be possibly nonvanishing.
 An example of such an observable is a sum of translated, local spin-flip operators, acting on an odd number of adjecent sites, 
\begin{align}
A_k=\sum_{x=1}^n \mathcal{S}^x(\underbrace{\sigma^x\otimes...\otimes\sigma^x}_{k}\otimes\one^{\otimes (n-k)}),
\end{align}
$k$ odd, which is obviously an observable of positive (or even) parity $\nu=+1$. We can now, for example, calculate the projection of $A_k$ onto the conserved quantities from $m$-dimensional irreducible semicyclic representations. Note that, since our observable has positive parity, the choice of particular semicyclic representation ($\Va_1,\Va_2$, or $\Va_3$) is irrelevant, since $(A_k,X_n(\varphi))=(A_k,PX_n(\varphi)P)$. The projections $(A_k,q_n(\varphi))$ can be left out of calculation of scalar products, since we are working in the thermodynamic limit and hence, using Cauchy-Schwarz inequality and quasilocality \eqref{qquasi} of local operator terms, we can estimate
\begin{align}
|(A_k,q_n(\varphi))|\;\le\;||A_k||_{\scriptscriptstyle HS}\,||q_n(\varphi)||_{\scriptscriptstyle HS}\;\le\; \zeta\, n\, e^{-\gamma\, n}.
\end{align}
The corresponding term in \eqref{projection} thus vanishes in the thermodynamic limit. Computation gives
\begin{align}
\lim_{n\rightarrow\infty}\frac{1}{n}(A_k,X_n(\varphi))&=(-1)^{(k-m)/2} \frac{g(s_1,\ldots,s_{m-1})}{2^k}\,[s_2s_3\cdots s_{m-1}]\,(\csc\varphi)^{k-2},\label{projectioncase}
\end{align}
where, again, $s_j=\sin(j\eta)$ and $g$ denotes some function of variables $s_1,\ldots,s_{m-1}$, accounting for the action of $(k-m)/2$ pairs of Lax operators $\mm{L}^+_0(\varphi)$, $\mm{L}^-_0(\varphi)$ that are, in addition to $m-2$ operators $\mm{L}^+_0(\varphi)$, present in the coefficients of $X_n(\varphi)$.\footnote{$m-2$ operators $\mm{L}^+_0(\varphi)$, coupling vectors $\ket{1}$ and $\bra{m-1}$, contribute the additional factor $s_2,\ldots s_{m-1}$.} As an example, let us calculate the explicit form of function $g$ in a particular case.
\begin{exm}
Let us take $k=9$ and $m=5$, $\eta=2\pi/5$. Non-zero terms in the projection $(A_k,X_n(\varphi))$ are supplied by terms of operators $q_k(\varphi)$ in $X_n(\varphi)$, consisting of three operators $\mm{L}_0^+(\varphi)$ and two pairs of operators
$\mm{L}^+_0(\varphi)$, $\mm{L}^-_0(\varphi)$. Using just relevant indices of these operators, we can write down nontrivial terms as
\begin{align}
\begin{aligned}
&\bra{4}++--+++\ket{1}\hspace{2cm}s_2s_3s_3s_4,\\
&\bra{4}+-+++-+\ket{1}\hspace{2cm}s_1s_2s_3s_4,\\
&\bra{4}+++-+-+\ket{1}\hspace{2cm}s_1s_2s_1s_2,\\
&\bra{4}++-++-+\ket{1}\hspace{2cm}s_1s_2s_2s_3,\\
&\bra{4}+++--++\ket{1}\hspace{2cm}s_1s_2s_2s_3,\\
&\bra{4}++-+-++\ket{1}\hspace{2cm}s_2s_3s_2s_3,\\
&\bra{4}+-++-++\ket{1}\hspace{2cm}s_2s_3s_3s_4,\\
&\bra{4}+-+-+++\ket{1}\hspace{2cm}s_3s_4s_3s_4.
\end{aligned}
\end{align}
Lax components \eqref{lax2} have been used. The common prefactor $s_2s_3s_4$, provided by the triple of $\mm{L}^+_0(\varphi)$ components, coupling $\ket{1}$ to $\bra{4}$, has been left out. The sum of these terms corresponds to a function $g$, used in Eq.~\eqref{projectioncase},
\begin{align}
g(s_1,s_2,s_3,s_4)=s_1^2s_2^2+s_2^2s_3^2+s_3^2s_4^2+2s_1s_2s_2s_3+2s_2s_3s_3s_4+s_1s_2s_3s_4.
\end{align}
From the form of Lax components, it can be deduced, that the function $g$, contributed by the action of $(k-m)/2$ pairs of operators $\mm{L}^+_0(\varphi)$ and $\mm{L}^-_0(\varphi)$, must always be a homogeneous polynomial of order $(k-m)/2$, in variables $s_1s_2,\ldots,s_{m-2}s_{m-1}$.
\end{exm}
Using Eqs.~\eqref{hsnorm}, \eqref{projection}, \eqref{fredholmeq}, \eqref{projectioncase} we see that, in order to calculate the Mazur-Suzuki lower bound, the solution of Fredholm equation
\begin{align}
\int_{\mathcal{D}_m} \d{^2\varphi'}\,\frac{\sin\varphi\,\sin\varphi'\,\sin(\varphi+\varphi')}{\sin(m(\varphi+\varphi'))}f(\varphi')=C\,(\csc\varphi)^{k-2},\label{fredholm}
\end{align}
for some constant $C$, is needed. At the moment we are not able to provide general analytic solutions $f(\varphi)$ to this equation, which however can easily be solved numerically.

As a possible physical realization, one may consider a quantum quench protocol from the Hamiltonian $H_\lambda= H_{XXZ} - \lambda A_k $ to $H_{XXZ}$ at $t=0$, restoring $U(1)$-symmetry of the $XXZ$ Hamiltonian (which is broken by $k-$spin operator $A_k$), starting with a thermal state $\rho(t < 0) = \rho_\beta = Z^{-1}\exp(-\beta H_\lambda)$.
For weak quenches (small $\lambda$), one can use linear response theory \cite{marcin}, to predict the after-quench steady state 
\begin{equation}
\rho_{\rm ss} = \rho_\beta (1 + \beta \lambda \bar{A_k}).
\end{equation}
Our generalized Drude weight $D_{A_k}$ thus yields a high-temperature Hilbert-Schmidt norm of all the magnetization non-conserving terms in the steady state, and can be bounded from below (\ref{eq:bound}) using the newly constructed quasi-local operators. This is an indicatetion that the latter have to be included in a complete generalized Gibbs ensemble of the $XXZ$ model at roots of unity.

\section{Conclusion}

The present paper extends the construction of quasilocal conservation laws for the anisotropic Heisenberg $XXZ$ model with periodic boundary conditions, developed in Refs. \cite{pro,pereira}, to the case of exotic representaiton structures provided by the quantum group \uq at roots of unity, namely the semicyclic representations. The conservation law property and commutativity are discussed in relation to the dimension of representation: only representations with dimension equal to the order of root of unity generate conserved quantities in involution. Newly constructed conserved quantities are quasilocal and do not conserve $z$-magnetization, i.e. they do not possess the $U(1)$ symmetry of the Hamiltonian. Interesting remaining problems and questions which should be investigated in future include: (i) physical applications of newly constructed quasilocal conserved charges, for example in relation to quench dynamics and generalized thermalizaiton, and (ii) analytically solving the Fredholm equation \eqref{fredholm}.

\section*{Acknowldegements}

This work has been supported by grants P1-0044, J1-5439 and N1-0025 of Slovenian Research Agency (ARRS). The authors thank A. Kl\"umper and F. G\" ohmann for helpful remarks, Christian Korff for enlightening discussion on the existence of intertwiners, and Enej Ilievski for useful comments regarding the manuscript.
TP also acknowledges Enej Ilievski for discussing initial ideas \cite{enejPHD} which gave birth to this work.

\end{document}